\documentstyle [12pt]{article}
\def\ZzZ{{\hbox{\rm Z\kern-.31em{Z}}}}
\def\CcC{{\hbox{\rm C\kern-.45em{\vrule height.67em width0.08em depth-
.04em \hskip.45em }}}}

\newcommand{\lab}{\label}

\newcommand{\bc}{\begin{center}}
\newcommand{\ec}{\end{center}}
\newcommand{\be}{\begin{equation}}
\newcommand{\ee}{\end{equation}}
\newcommand{\bea}{\begin{eqnarray}}
\newcommand{\eea}{\end{eqnarray}}
\newcommand{\bs}{\begin{subequations}}
\newcommand{\es}{\end{subequations}}
\newcommand{\beq}{\begin{eqalignno}}
\newcommand{\eeq}{\end{eqalignno}}

\def\lab{\label}

\begin{document}

\thispagestyle{empty}

\vspace{2.0cm}
\bc
\Huge{Thermo Field Dynamics and quantum algebras}
\vspace{1.5cm}

\large{ E.Celeghini$^{\dag}$, S.De Martino$^{*}$, S.De Siena$^{*}$,
A.Iorio$^{**}$, M.Rasetti$^{+}$ and G.Vitiello$^{*}$ }\\
\vspace{1.0cm}
{\it ${}^{\dag}$Dipartimento di Fisica, Universit\`a di Firenze, and 
INFN-Firenze, I-50125 Firenze, Italy}\\
{\it ${}^{*}$Dipartimento di Fisica, Universit\`a di Salerno, and
INFN-Salerno, I-84100 Salerno, Italy}\\
{\it ${}^{**}$School of Mathematics, Trinity College, Dublin, Ireland }\\
{\it ${}^{+}$Dipartimento di Fisica and Unit\`a INFM, Politecnico di Torino,
I-10129 Torino, Italy} \\
\normalsize

\ec
\normalsize




\bigskip
\bigskip
\bigskip

 {\bf Abstract} 

The algebraic structure of Thermo Field Dynamics lies in the $q$-deformation
of the algebra of creation and annihilation operators. Doubling of the degrees
of freedom, tilde-conjugation rules, and Bogoliubov transformation for bosons
and fermions are recognized as algebraic properties of $h_{q}(1)$ and of
$h_{q}(1|1)$, respectively.

\vskip 1.75truecm \noindent

 



\newpage
\setcounter{page}{1}
\setcounter{equation}{0}
\section*{1 Introduction}

One central ingredient of Hopf algebras~\cite{CP} is the operator doubling
implied by the coalgebra. The coproduct operation is indeed a map ${\Delta}: 
{\cal A}\to {\cal A}\otimes {\cal A}$ which duplicates the algebra. Lie-Hopf
algebras are commonly used in the familiar addition of energy, momentum and
angular momentum.

On the other hand, the doubling of the degrees of freedom turns out to be the
central ingredient also in thermal field theories in the formalism of Thermo
Field Dynamics (TFD)~\cite{TFD} which has been recognized~\cite{Oj} to be
strictly related with the $C^{*}$ algebras formalism~\cite{BR}.  In this
paper we show that this is not a merely formal feature, but that the natural
TFD algebra is indeed the Hopf algebra of creation and annihilation operators.
Preliminary results in such a direction were presented in~\cite{DDV} and the
strict connection between TFD and bialgebras has been also discussed in~
\cite{SK}.

To be more specific we show that the set of the algebraic rules called the
$''$tilde-conjugation rules$\, ''$, axiomatically introduced in TFD (and in
its $C^{*}$-algebraic formulation), as well as the Bogoliubov transformation
and its generator follow from basic and simple properties of quantum Hopf
algebras. In particular, the deformed Weyl-Heisenberg algebra $h_{q}(1)$
describes the TFD for bosons while the quantum deformation  $h_{q}(1|1)$ of
the customary superalgebra of fermions describes the TFD for fermions. 

In sec. 2 we briefly introduce $h_q(1)$ and $h_q(1|1)$. In sec. 3 we present
our main result: from the Hopf algebra properties we derive tilde conjugation
rules and Bogoliubov transformations with their generator. Sec. 4 is devoted
to further discussions.

\setcounter{equation}{0}
\section*{2 The creation and annihilation operator algebras }

The bosonic algebra $h(1)$ is  generated by the set of operators $\{ a,
a^{\dagger},H,N \}$ with commutation relations:
\be
[\, a\, ,\, a^{\dagger} \, ] = \ 2H \, , \quad\; 
[\, \ N \, ,\, a \, ] = - a \, , \quad\; [\, \ N \, ,\, a^{\dagger} \, ] = 
a^{\dagger} \, , \quad\; [\, \ H \, ,\, \bullet \, ] = 0 \, . 
\lab{p22}
\ee
$H$ is a central operator, constant in each representation. The Casimir
operator is given by ${\cal C} = 2NH -a^{\dagger}a$.
~$h(1)$ is an Hopf algebra and is therefore equipped with the coproduct
operation, defined by
\be
\Delta a = a \otimes {\bf 1} + {\bf 1} \otimes a \equiv a_1 + a_2 ~,~~~
~\Delta a^{\dagger} = a^{\dagger} \otimes {\bf 1} + {\bf 1} \otimes 
a^{\dagger} \equiv a_1^{\dagger} + a_2^{\dagger} ~,  
\lab{p23}\ee
\be
\Delta H = H \otimes {\bf 1} + {\bf 1} \otimes H  \equiv H_{1} + 
 H_{2}~, ~~~\Delta N = N \otimes {\bf 1} + {\bf 1} \otimes N \equiv  N_{1} +
 N_{2} ~.  \lab{p24}
 \ee

The physical meaning of the coproduct is that it provides the prescription
for operating on two modes.  One example of coproduct is the familiar
operation performed with the $''$addition$\, ''$ of the angular momentum
$J^{\alpha}$, ${\alpha} = 1,2,3$, of two particles:
$\Delta J^{\alpha} = J^{\alpha}  \otimes {\bf 1} + {\bf 1} \otimes
J^{\alpha} \equiv J^{\alpha}_1 + J^{\alpha}_2,~ J^{\alpha} \in su(2)$.

The $q$-deformation of $h(1)$, $h_{q}(1)$, with deformation parameter $q$, is:
\be
[\, a_{q}\, ,\, a_{q}^\dagger \, ] = \ [2H]_{q} \, , \quad\;
[\, \ N \, \, , \, a_{q}\, ] = - a_{q} \, , \quad\;  [ \, \ N \, , \, 
a_{q}^\dagger \,] = a_{q}^\dagger , \quad\; [\, \ H \, \, , \,  \bullet \,
] = 0  \, , \lab{p26}
\ee
where $N_{q} \equiv N$ and $H_{q} \equiv H$.  The Casimir operator ${\cal
C}_{q}$ is given by ${\cal C}_{q} = N[2H]_{q} -a_{q}^{\dagger}a_{q}$, where
$\displaystyle{[x]_{q} = {{q^{x} - q^{-x}} \over {q - q^{-1}}}}$. Also 
$h_{q}(1)$ is an Hopf algebra and its coproduct is defined by
\be
\Delta a_{q} = a_{q} \otimes {q^{H}} + { q^{-H}} \otimes a_{q} 
\, , \quad\quad \Delta a_{q}^{\dagger} = a_{q}^{\dagger} \otimes {q^H} +
{q^{-H}} \otimes a_{q}^{\dagger} ~, ~\lab{p28}
\ee
\be
\Delta H = H \otimes {\bf 1} + {\bf 1} \otimes H ~,~~~~ \Delta N
= N \otimes {\bf 1} + {\bf 1} \otimes N ~, \lab{p29}
\ee
whose algebra of course is isomorphic with (\ref{p26}): ~$ [ \Delta a_{q} ,
\Delta a_{q}^{\dagger} ] = [2 {\Delta} H]_{q}$ , etc. .

We denote by ${\cal F}_{1}$  the single mode Fock space, i.e. the fundamental
representation $H = 1/2$, ${\cal C} = 0$. In such a representation $h(1)$ and
$h_{q}(1)$ coincide as it happens for $su(2)$ and $su_{q}(2)$ for the
spin-$\frac{1}{2}$ representation. The differences appear in the coproduct
and in the higher spin representations.

In the case of fermions the $\ZzZ_2$-graded algebra $h(1|1)$ is generated by
the relations 
\be
\{ \, a\, ,\, a^{\dagger} \, \} = \ 2H \, , \quad\;
[\, \ N \, ,\, a \, ] = - a \, , \quad\; [\, \ N \, ,\, a^{\dagger} \, ] = 
a^{\dagger} , \quad\; [\, \ H \, ,\, \bullet \, ] = 0  ~. 
\lab{p211}
\ee

Also $h(1|1)$ is an Hopf algebra (actually a Hopf superalgebra), equipped with
the same coproduct operations defined in (\ref{p23})-(\ref{p24}). The deformed
algebra $h_{q}(1|1)$ relations are as those in (\ref{p26}), with the first
commutator replaced by the anti-commutator and its coproduct is defined as in
(\ref{p28})-(\ref{p29}).

As customary, we require that $a$ and $a^{\dag}$, and $a_{q}$ and
${a_{q}}^{\dag}$,  are adjoint operators. This implies that $q$ can only be
real or of modulus one.

In the two mode Fock space ${\cal F}_{2} = {\cal F}_{1} \otimes {\cal F}_{1}$,
for $|q|=1$, the hermitian conjugation of the coproduct must be supplemented
by the inversion of the two spaces for consistency with the coproduct
isomorphism.

Summarizing we can write for both bosons and fermions on ${\cal F}_{2} =
{\cal F}_{1} \otimes {\cal F}_{1}$:
\be
\Delta a =  a_1 + a_2 ~,~~~ ~\Delta a^{\dagger} = a_1^{\dagger} +
a_2^{\dagger} ~, \lab{p212}
\ee
\be
\Delta a_{q} =  a_1 q^{1/2} + q^{-1/2} a_2 ~,~~~
~\Delta a_{q}^{\dagger} = a_1^{\dagger} q^{1/2}  +q^{-1/2}  a_2^{\dagger} ~,  
\lab{p213}
\ee
\be
\Delta H = 1 , ~~~\Delta N =  N_{1} +
 N_{2} ~.  \lab{pdelta} \lab{p214}
\ee

We observe that ${[a_i , a_j ]}_{\mp} = {[a_i , a_{j}^{\dagger} ]}_{\mp} = 0 ,
~ i \neq j $. Here, and in the following, $[~ ,~ ]_{-}$ and $[~ ,~ ]_{+}$
denote, respectively, commutators and anticommutators. 

\section*{3 Coproduct and the Bogoliubov transformation}

In this section we derive our main result, namely that the algebraic structure 
on which the TFD formalism is based is naturally provided by the Hopf algebras
$h_q(1)$ and $h_q(1|1)$.

It is convenient to start recalling the so-called $''$tilde-conjugation
rules$\, ''$ which are defined in TFD. For any two bosonic (respectively,
fermionic) operators $A$ and $B$ and any two $c$-numbers ${\alpha}$ and
${\beta}$ the tilde-conjugation rules of TFD are postulated to be the
following~\cite{TFD}:
\be
(AB)^{\tilde {}} = {\tilde {A}}{\tilde {B}}~, \lab{p31}
\ee
\be
({\alpha}A +{\beta}B)^{\tilde {}} = {\alpha}^{*}{\tilde {A}} +
{\beta}^{*}{\tilde {B}}~, \lab{p32}
\ee
\be
(A^{\dagger})^{\tilde {}} = {\tilde {A}}^{\dagger}~,
\lab{p33}
\ee
\be
({\tilde A})^{\tilde {}} = A~.
\lab{p34}
\ee
According to (\ref{p31}) the tilde-conjugation does not  change the order
among operators. Furthermore, it is required that tilde and non-tilde
operators are mutually commuting (or anti-commuting) operators and that the
thermal vacuum $|0(\beta)>$ is invariant under tilde-conjugation:
\be
[ A , {\tilde B}]_{\mp} = 0 = [ A , {\tilde B}^{\dagger} ]_{\mp} ~, \lab{p35}
\ee
\be
{|0(\beta)>}^{\tilde {}} = |0(\beta)> ~. \lab{p36}
\ee
In order to use a compact notation it is useful to introduce the parity label
$\sigma$ defined by $\sqrt{\sigma} \equiv +1$ for bosons and $\sqrt{\sigma}
\equiv + i$ for fermions.  We shall therefore simply write commutators
as $[ A,B ]_{-\sigma} \doteq AB-\sigma BA$, and $(1 \otimes A )(B \otimes 1)
\equiv\sigma (B \otimes 1)(1 \otimes A)$,without further specification of
whether $A$ and $B$ (which are equal to $a$, $a^{\dagger}$ in all possible
ways) are fermions or bosons.

As it is well known, the central point in the TFD formalism is the possibility
to express the statistical average $<A>$ of an observable $A$ as the
expectation value in the temperature dependent vacuum $|0(\beta)>$:
\be
<A>~ \equiv~ {Tr[A~e^{-{\beta}{\cal H}}] \over{Tr[e^{-{\beta}{\cal H}}]}}~ =~
<0(\beta)|A|0(\beta)> \; , \lab{p361}
\ee
and this requires the introduction of the tilde-degrees of freedom~\cite{TFD}.

Our first statement is that the doubling of the degrees of freedom on which 
the TFD formalism is based finds its natural realization in the coproduct map.
Upon identifying from now on $a_{1} \equiv a$, $a_{1}^{\dagger} \equiv
a^{\dagger}$, one easily checks that the TFD tilde-operators (consistent with
(\ref{p31}) -- (\ref{p35})) are straightforwardly recovered by setting $a_{2}
\equiv {\tilde a}$ , $a_{2}^{\dagger} \equiv {\tilde a}^{\dagger}$. In other
words, according to such identification, it is the action of the ~$1
\leftrightarrow 2$~ permutation $\pi$: $\pi a_{i} = a_{j} ,~~ i\neq j ,~~ i,j
=1,2$, that defines the operation of $''$tilde-conjugation$\, ''$:
\be
\pi a_{1} = \pi (a \otimes {\bf 1} ) = {\bf 1} \otimes a = a_{2}
\equiv {\tilde a} \equiv  {( a )^{\tilde {} }}
\lab{p37}
\ee
\be
\pi a_{2} = \pi ({\bf 1} \otimes a) = a \otimes {\bf 1} = a_{1} \equiv a
\equiv {({\tilde a})^{\tilde {} }}  ~ . \lab{p38}
\ee

In particular, being the $\pi$ permutation involutive, also tilde-conjugation
turns out to be involutive, as in fact required by the rule (\ref{p34}).
Notice that, as $(\pi a_{i})^{\dagger} = \pi ({a_i}^{\dagger})$, it is also
${ ( { (a_i)^{\tilde {}} }~ )^{\dagger} } = ((a_i)^{\dagger})^{\tilde {} }$, 
i.e. tilde-conjugation commutes with hermitian conjugation.
Furthermore, from  (\ref{p37})-(\ref{p38}), we have
\be
{(ab)^{\tilde {} }} = {[(a \otimes {\bf 1})(b \otimes {\bf 1})]^{\tilde {}}} 
= {(ab \otimes {\bf 1})^{\tilde {} }} = {\bf 1} \otimes ab = ({\bf 1} \otimes
a)({\bf 1} \otimes b) = {\tilde a}{\tilde b} \; . \lab{p39}
\ee
Rules (\ref{p33}) and (\ref{p31}) are thus obtained. (\ref{p35}) is insured
by the $\sigma$-commutativity of $a_{1}$ and $a_{2}$. The vacuum of TFD,~
$|0(\beta)>$, ~is a condensed state of equal number of tilde and non-tilde
particles~\cite{TFD}, thus (\ref{p36}) requires no further conditions: eqs.
(\ref{p37})-(\ref{p38}) are sufficient to show that the rule (\ref{p36}) is
satisfied.

Let us now consider the following operators:
\be
A_{q} \equiv { { {\Delta} a_{q}} \over {\sqrt{[2]_{q}} }} =
{1\over\sqrt{[2]_{q}}} (e^{{\sqrt{\sigma}}\theta} a +
e^{-{\sqrt{\sigma}}\theta} \tilde{a} ) ~, \lab{p310}
\ee
\be
B_{q} \equiv { 1 \over {\sqrt{[2]_{q}}}{\sqrt{\sigma}} } {\delta \over
{\delta \theta}} {\Delta} a_{q} = {2q \over \sqrt{[2]_{q}}}{{\delta}\over
{\delta q}} \Delta a_{q} = {1\over\sqrt{[2]_{q}}} (e^{{\sqrt{\sigma}}\theta}
a -e^{-{\sqrt{\sigma}}\theta} \tilde{a} ) \; ,
\lab{p311}\ee
and h.c., with $q = q(\theta) \equiv e^{{\sqrt{\sigma}2\theta}}$ . Notice 
that
\be
{{\delta}\over{{\sqrt{\sigma}}\delta \theta}}
({{\delta}\over{{\sqrt{\sigma}}\delta \theta}}
\Delta a_{q} )= \Delta a_{q} ~. \lab{p312}
\ee
The commutation and anti-commutation relations are
\be
[ A_{q} , A_{q}^{\dagger} ]_{- \sigma} = 1 ~, ~~[ B_{q} ,
B_{q}^{\dagger} ]_{- \sigma} = 1 ~,~~ [ A_{q} , B_{q} ]_{- \sigma} = 0~,
 ~[ A_{q} , B_{q}^{\dagger} ]_{- \sigma} =
{1 \over{\sqrt{\sigma}}}~ {\rm tanh} ~{{\sqrt{\sigma}}2\theta}~,~~
\lab{p313}
\ee
whereas all other $\sigma$-commutators equal zero. A set of commuting
operators with canonical commutation relations is given for bosons ($\sigma
=1$) by
\be
A(\theta) \equiv {{\sqrt{[2]_{q}}}  \over 2{\sqrt2}} [ A_{q(\theta)} +
A_{q(- \theta)} - B_{q(\theta)}^{\dagger}  +  B_{q(- \theta)}^{\dagger}] ~,
\lab{p314}
\ee
\be
B(\theta) \equiv {{\sqrt{[2]_{q}}}  \over 2{\sqrt 2}}[B_{q(\theta)} +
B_{q(- \theta)} - A_{q(\theta)}^{\dagger}  +  A_{q(- \theta)}^{\dagger} ] ~.
\lab{p315}
\ee
and h.c. Analogously, for fermions ($\sigma = -1$), anti-commuting operators
are given by:
\be
A(\theta) \equiv {{\sqrt{[2]_{q}}}  \over 2{\sqrt2}} [ A_{q(\theta)} +
A_{q(- \theta)} + A_{q(\theta)}^{\dagger}  -  A_{q(- \theta)}^{\dagger}] ~,
\lab{p3141}
\ee
\be
B(\theta) \equiv {{\sqrt{[2]_{q}}}  \over 2{\sqrt 2}}[B_{q(\theta)} +
B_{q(- \theta)} - B_{q(\theta)}^{\dagger}  +  B_{q(- \theta)}^{\dagger} ]~.
\lab{p3151}
\ee
and h.c.  One has 
\be
[ A(\theta) , A^{\dagger}(\theta) ]_{- \sigma} = 1 ~, ~~[ B(\theta) ,
B^{\dagger}(\theta) ]_{- \sigma} = 1 ~,~~ [ A(\theta) , B^{\dagger}(\theta)
]_{- \sigma} = 0 \; , \lab{p316}
\ee
and all other $\sigma$-commutators equal to zero. Of course, here it is 
understood that the operators given by eqs. (\ref{p314})-(\ref{p315})
commute and the operators given by eqs. (\ref{p3141})-(\ref{p3151}) 
anticommute.

We can also write, for both bosons and fermions,
\be
A(\theta) = {1 \over {\sqrt 2}}(a(\theta) + {\tilde a}(\theta)) ~,~~~
B(\theta) = {1 \over {\sqrt 2}}(a( \theta ) - {\tilde a}(\theta )) ~,
\lab{p320}
\ee
with
\bea
a(\theta) = \frac{1}{\sqrt{2}} \left ( A(\theta ) + B(\theta )\right )
&=& a ~{\rm cosh} ~{\sqrt\sigma}\theta - {\tilde a}^{\dagger}
~{\rm sinh} ~{\sqrt\sigma}\theta ~~, \nonumber \\ 
{\tilde a}(\theta) = \frac{1}{\sqrt{2}} \left ( A(\theta ) - B(\theta )
\right ) &=& {\tilde a} ~{\rm cosh} ~{\sqrt\sigma}\theta -
{\sigma} a^{\dagger} ~{\rm sinh} ~{\sqrt\sigma}\theta ~, \lab{p321}
\eea
\be
[ a(\theta) , a^{\dagger}(\theta) ]_{- \sigma} = 1 ~, ~~[ {\tilde a}(\theta) , 
{\tilde a}^{\dagger}(\theta) ]_{- \sigma} = 1 ~.~~
\lab{p322}
\ee
All other $\sigma$-commutators are equal to zero and $a(\theta)$ and ${\tilde
a}(\theta)$ $\sigma$-commute among themselves. Eqs. (\ref{p321}) are nothing
but the Bogoliubov transformations for the $(a, {\tilde a})$ pair into a new
set of creation, annihilation operators.  In other words, eqs. (\ref{p321}),
(\ref{p322}) show that the Bogoliubov-transformed operators $a(\theta)$ and
${\tilde a}(\theta)$ are linear combinations of the coproduct operators
defined in terms of the deformation parameter $q(\theta )$ and of their
${\theta}$-derivatives; namely the Bogoliubov transformation is implemented
in differential form (in $\theta$) as
\bea 
a(\theta ) &=& \frac{1}{4} 
\left ( 1 + \frac{1}{\sqrt{\sigma}} \, \frac{\delta}{\delta \theta} \right )
\;\;\; \Delta\left [ a_q + a_{q^{-1}} 
- ( {a_q}^{\dagger} - {a_{q^{-1}}}^{\dagger} )
\right ] \; \nonumber \\
&=& \frac{1}{\sqrt{2}}\left [
e^{  \alpha ( 1 + \frac{1}{\sqrt{\sigma}} \, \frac{\delta}{\delta \theta})} -
e^{- \alpha ( 1 + \frac{1}{\sqrt{\sigma}} \, \frac{\delta}{\delta \theta})}
\right ] \;\;\;
\Delta\left [ a_q + a_{q^{-1}} 
- ( {a_q}^{\dagger} - {a_{q^{-1}}}^{\dagger} )
\right ] \;\;\;\;
\lab{p3221}
\eea
\bea 
\tilde{a}(\theta ) &=& \frac{1}{4} \left ( 1 - \frac{1}{\sqrt{\sigma}} \,
\frac{\delta}{\delta \theta} \right ) 
\;\;\; \Delta\left [ a_q + a_{q^{-1}} + \sigma ( {a_q}^{\dagger} - 
{a_{q^{-1}}}^{\dagger})\right ] \;  \nonumber \\
&=& \frac{1}{\sqrt{2}} \left [ 
e^{  \alpha ( 1 - \frac{1}{\sqrt{\sigma}} \, \frac{\delta}{\delta \theta})} -
e^{- \alpha ( 1 - \frac{1}{\sqrt{\sigma}} \, \frac{\delta}{\delta \theta})}
\right ] \;\;\; \Delta\left [ a_q + a_{q^{-1}} + \sigma ( {a_q}^{\dagger} - 
{a_{q^{-1}}}^{\dagger})\right ] \;\;\;\;
\lab{p3222}
\eea
where $\alpha = \frac{1}{4} \log 2$ .

Note that inspection of Eq.~(\ref{p321}) in the fermion case shows that
$\sqrt{\sigma} (= i)$ changes sign under tilde-conjugation. This is related
to the antilinearity property of tilde-conjugation, which we shall discuss in
more detail below.

Next, we observe that the ${\theta}$-derivative, namely the derivative with
respect to the $q$-deformation parameter, can be represented in terms of
commutators of $a(\theta)$ (or of ${\tilde a}(\theta)$) with the generator
${\cal G}$ of the Bogoliubov transformation (\ref{p321}). 

{}From (\ref{p321}) we see that ${\cal G}$ is given by
\be
{\cal G} \equiv -i~{\sqrt{\sigma}}~(a^{\dagger} {\tilde a}^{\dagger}
 - a {\tilde a}) ~, ~~ \lab{p323}
\ee
\be
a(\theta) = \exp(i{\theta}{\cal G})~ a~ \exp(-i{\theta}{\cal G})~,~{\tilde
a}(\theta) = \exp(i{\theta}{\cal G}) ~{\tilde a}~ \exp(-i{\theta}{\cal G}) ~.
\lab{p324}
\ee
Notice that, because of(\ref{p321}) (and (\ref{p312})), 
\bea
{\delta \over {\delta \theta}} a(\theta) = - {\sqrt{\sigma}}{\tilde 
a}^{\dagger} (\theta) ~ &,& ~ {\delta \over {\delta \theta}} {\tilde a}
(\theta) = - ({\sqrt{\sigma}}a (\theta))^{\dagger}  ~, \nonumber \\ 
 {\delta^{2} \over {\delta \theta}^{2}} a(\theta) = {\sigma}a(\theta) ~ &,& ~
 {\delta^{2} \over {\delta \theta}^{2}} {\tilde a}(\theta) =
 {\sigma}{\tilde a}(\theta) ~,
\lab{p325}
\eea
and h.c.~. The relation between the ${\theta}$-derivative and ${\cal G}$ is
then of the form:
\be
- i{\delta \over {\delta \theta}} a(\theta) =
[{\cal G}, a(\theta)] ~,~~~
- i{\delta \over {\delta \theta}} {\tilde a}(\theta) =
[{\cal G}, {\tilde a}(\theta)] ~,
\lab{p326}\ee
and h.c.~.
For a fixed value $\bar{\theta}$, we have
\be
\exp(i{\bar{\theta}} p_{\theta}) ~a(\theta) =
\exp(i{\bar{\theta}}{\cal G}) ~a(\theta)~
\exp(-i{\bar{\theta}}{\cal G}) = a( \theta + {\bar {\theta}} )
~,
\lab{p327}\ee
and similar equations for ${\tilde a}(\theta)$.

In eq.(\ref{p327}) we have used the definition $\displaystyle{p_{\theta}
\equiv -i{\delta \over {\delta \theta}}}$ which can be regarded as the
momentum operator $''$conjugate$\, ''$ to the $''$thermal degree of
freedom$\, ''$ $\theta$ (in TFD of course $\theta \equiv \theta (\beta)$,
with $\beta$ the inverse temperature). The notion of thermal degree of freedom
\cite{Banf} thus acquires formal definiteness in the sense of the canonical
formalism. It is interesting to observe that derivative with respect to the
$q$-deformation parameter is actually a derivative with respect to the system
temperature T. This may shed some light on the r\^ole of $q$-deformation in
thermal field theories for non-equilibrium systems and phase transitions. We
shall comment more on this point in the following section.

Eqs. (\ref{p325}) show that the tilde-conjugation may be represented by the
$\theta$-derivative in the following way:
\be
\left ( -{\delta\over{{\sqrt\sigma}\delta\theta}} a(\theta)\right )^{\dagger}
\Biggr |_{{\theta} = 0} = {\tilde a} = {( a )^{\tilde {}}} ~, ~~~~
\left ({1\over \sigma}{\delta^{2} \over {\delta \theta}^{2}} a(\theta)
\right ) \Biggr |_{{\theta} = 0} = a = {( {\tilde a})^{\tilde {}}}  ~, 
\lab{p329}
\ee
and h.c.. We also have
\be
\left ( -{\delta \over{{\sqrt\sigma}\delta \theta} } {\alpha} 
a(\theta) \right )^{\dagger} \Biggr |_{{\theta} = 0} = \left ( -{\delta \over
{({\sqrt\sigma})^*\delta \theta}} {\alpha}^{*} a(\theta)^{\dagger}\right )
\Biggr |_{{\theta} = 0} = {\alpha}^{*}{\tilde a} (\theta)\bigr |_{{\theta} =
0} = {\alpha}^{*}{\tilde a} ~, ~~ \lab{p330}
\ee
where ${\alpha}$ denotes any $c$-number. Thus tilde-conjugation is antilinear,
as in fact it is required in TFD (rule (\ref{p32})).

We finally observe that, from eqs. (\ref{p325}) (and (\ref{p39})), we have
\be
\left ( -{\delta \over {{\sqrt\sigma}\delta \theta}}a(\theta)\right
)^{\dagger} \, \left ( -{\delta \over {{\sqrt\sigma}\delta \theta}} b(\theta)
\right )^{\dagger}  = {\tilde a}(\theta){\tilde b}(\theta) = {(a(\theta)b(
\theta))^{\tilde {}}} ~, \lab{p331}
\ee
for any ${\theta}$, again in agreement with TFD.

Under tilde-conjugation $a {\tilde a}$ goes into ${\tilde a} a = {\sigma} a
{\tilde a}$. From this we note that $a {\tilde a}$ is not tilde invariant.
Since $\sqrt{\sigma} = i$ when $\sigma = -1$, and $i$ changes sign under
tilde-conjugation, we see that instead ${\sqrt{\sigma}} a {\tilde a}$ is
tilde invariant~\cite{Banf}. We also remark that $\Delta a_q$ and $\Delta
{a_q}^\dagger $ are tilde-invariant in the fermion case.

In conclusion, the doubling of the degrees of freedom and  the 
tilde-conjugation rules, which in TFD are postulated, are shown to be
immediate consequences of the coalgeebra structure (essentially the coproduct
map), of the $\pi$ permutation and of the derivative with respect to the
deformation parameter in a $q$-algebraic frame.  Moreover, in the $h_{q}(1)$
and $h_{q}(1|1)$ coalgebras, TFD appears also equipped with a set of
canonically conjugate $''$thermal$\, ''$ variables
$({\theta} , p_{\theta})$.

\section*{4 Inequivalent representations and the deformation parameter}

We note that in the boson case
$J_1 \equiv {1\over 2}(a^{\dagger}{\tilde a}^{\dagger}  + a{\tilde a})$ 
together with $J_2 \equiv {1\over 2}{\cal G}$ and $J_{3} \equiv {1 \over 2}
( N  + {\tilde N} + 1)$ 
close an algebra $su(1,1)$. Moreover, ${\delta \over {\delta \theta}}(N(
\theta) - {\tilde N}(\theta)) = 0$ , with $(N(\theta) - {\tilde N}(\theta))
\equiv (a^{\dagger}(\theta)a(\theta) - {\tilde a}^{\dagger}(\theta){\tilde a}
(\theta))$, consistently with the fact that ${1 \over 4}(N - {\tilde N})^{2}$
is the $su(1,1)$ Casimir operator.

In the fermion case  $ J_{1} \equiv {1\over 2}{\cal G}$, $J_{2}  \equiv 
{1\over 2}(a^{\dagger}{\tilde a}^{\dagger}  + a{\tilde a})$ and $J_{3} 
\equiv {1 \over 2}( N  + {\tilde N} - 1)$ close an algebra $su(2)$. Also in
this case ${\delta \over {\delta\theta}}(N(\theta) - {\tilde N}(\theta))
= 0$ , with $(N(\theta) - {\tilde N}(\theta)) \equiv (a^{\dagger}(\theta)a(
\theta ) - {\tilde a}^{\dagger}(\theta){\tilde a}(\theta))$, again
consistently with the fact that ${1 \over 4}(N - {\tilde N})^{2}$ is related
to the $su(2)$ Casimir operator.

The $su(1,1)$ algebra and the $su(2)$ algebra, which are the boson and 
the fermion TFD algebras, are thus described as well in terms of operators of
$h_{q}(1)$ and $h_{q}(1|1)$.

The vacuum state for $a(\theta)$ and ${\tilde a}(\theta)$ is formally given
(at finite volume) by
\be
|0(\theta)> ~=~ \exp ~(i{\theta}{\cal G})~ ~|0,0> ~=~ \sum_{n} c_{n} (\theta
)\, |n,n> ~, \lab{41}
\ee
with $n= 0,..\infty$ for bosons and $n = 0,1$ for fermions, and it 
appears therefore to be an $SU(1,1)$ or $SU(2)$ generalized coherent 
state~\cite{Pe}, respectively for bosons or for fermions.

In the infinite volume limit $|0(\theta)>$ becomes orthogonal to $|0,0>$
and we have that the whole Hilbert space $\{|0(\theta)>\}$, constructed by
operating on $|0(\theta)>$ with $a^{\dagger}(\theta)$ and ${\tilde
a}^{\dagger}(\theta)$, is asymptotically orthogonal to the space generated
over $\{|0,0>\}$. In general, for each value of the deformation parameter,
i.e. $\displaystyle{\theta = {1 \over {2 \sqrt{\sigma}}} ~{\rm ln}~q}$,
we obtain in the infinite volume limit a representation of the canonical
commutation relations unitarily inequivalent to the others, associated with 
different values of $\theta$. In other words, the deformation parameter
acts as a label for the inequivalent representations, consistently with a 
result already obtained elsewhere~\cite{IV}. In the TFD case $\theta = \theta
(\beta)$ and the physically relevant label is thus the temperature. The state
$|0(\theta)>$ ~is of course the thermal vacuum and the tilde-conjugation rule
(\ref{p36}) holds true together with ~$(N - {\tilde N})|0(\theta)> ~=~ 0$,
which is the equilibrium thermal state condition in TFD.

It is remarkable that the "conjugate thermal momentum" $p_{\theta}$ 
generates transitions among inequivalent (in the infinite volume limit) 
representations: $\exp (i{\bar \theta}p_{\theta})~ ~|0(\theta)> = |0(\theta +
{\bar \theta})>$.

In this connection let us observe that variation in time of the deformation
parameter is related with the so-called heat-term in dissipative systems. In
such a case, in fact, ${\theta} = {\theta}(t)$ (namely we have time-dependent
Bogoliubov transformations), so that the Heisenberg equation for $a(t,{\theta}
(t))$ is
\bea 
-i{\dot a}(t,{\theta}(t)) &=& -i{\delta \over {\delta t}} a(t,{\theta}(t))
-i{{\delta \theta} \over {\delta t}}~ {\delta  \over {\delta \theta}}a(t,
{\theta}(t)) = \nonumber \\
\left [ H , a(t,{\theta}(t)) \right ] +
{{\delta \theta} \over {\delta t}}~ [{\cal G}, a(t,{\theta}(t)) ] &=& 
\left [ H + Q , ~a(t,{\theta}(t)) \right ] ~, \lab{42}
\eea
where $\displaystyle{Q \equiv {{\delta \theta} \over {\delta t}} {\cal G}}$
denotes the heat-term~\cite{DDV}, \cite{Banf}, and $H$ is the hamiltonian
(responsible for the time variation in the explicit time dependence of
$a(t,{\theta}(t))$).  $H + Q$ is therefore to be identified rather with the
free energy~\cite{TFD}, \cite{DDV}, \cite{CRV}. When, as usual in TFD,
$H|0(\theta)>=0$, the time variation of the state $|0(\theta)>$ is given by 
\be
-i{\delta \over {\delta t}}|0(\theta)> = {i \over 2}{{\delta \theta} \over
{\delta t}}~\left ({\delta  \over {\delta \theta}} {\cal S}(\theta)
\right )~|0(\theta)> ~.
\lab{43}
\ee
Here ${\cal S}(\theta)$ denotes the entropy operator~\cite{TFD}, \cite{CRV}:
\be
{\cal S}(\theta) = - (a^{\dagger}a~{\rm ln~{\sigma}~ sinh}^{2}{{\sqrt{\sigma}}
\theta} - {\sigma}~ a a^{\dagger}~{\rm ln~cosh}^{2} {{\sqrt{\sigma}}\theta}) \; .
\lab{44}
\ee

We thus conclude that variations in time of the deformation parameter actually
involve dissipation.

Finally, when the proper field description is taken into account, $a$ and
${\tilde a}$ carry dependence on the momentum ${\bf k}$ and, as customary in
QFT (and in TFD), one should deal with the algebras $\displaystyle{
\bigoplus_{\bf k} h_{\bf k}(1)}$ and $\displaystyle{\bigoplus_{\bf k} h_{\bf
k}(1|1)}$.  In TFD this leads to expect that one should have ${\bf
k}$-dependence also for ${\theta}$. The Bogoliubov transformation analogously,
thought of as inner automorphism of the algebra $su(1,1)_{\bf k}$ (or
$su(2)_{\bf k}$), allows us to claim that one is globally dealing with
$\displaystyle{\bigoplus_{\bf k} su(1,1)_{\bf k}}$ (or $\displaystyle{
\bigoplus_{\bf k} su(2)_{\bf k}}$). Therefore we are lead to consider
${\bf k}$-dependence also for the deformation parameter, i.e. to consider 
$h_{q(\bf k)}(1)$ (or $h_{q(\bf k)}(1|1)$ ). In such a way the conclusions
presented in the former part of the paper can be extended to the case of many
degrees of freedom.

\vfill\eject
\newpage


\end{document}